\documentclass[journal]{IEEEtran}
\ifCLASSINFOpdf
\else
\fi
\usepackage{graphicx}
\usepackage{booktabs}
\usepackage{amsmath,array,graphicx}
\usepackage{kantlipsum}
\usepackage[noadjust]{cite}
\usepackage{filecontents}
\usepackage{silence}
\usepackage{caption}
\usepackage{float}
\usepackage[latin1]{inputenc}
\usepackage{tikz}
\usepackage[hyphens]{url}
\usepackage{tkz-graph}
\usepackage{float}
\restylefloat{table}
\usepackage{placeins}
\usetikzlibrary{shapes,arrows}
\usetikzlibrary{calc}
\usepackage{amssymb} 
\usepackage{blkarray}
\usepackage{siunitx,etoolbox}
\usepackage{tabularx}
\usepackage[tableposition = top]{caption}
\usepackage{nicematrix}
\usepackage{multirow}
\usepackage[hyphenbreaks]{breakurl}
\numberwithin{equation}{section}
\usepackage{adjustbox}
\usepackage{smartdiagram}
\usepackage{algpseudocode,algorithm,algorithmicx}  
\usepackage[normalem]{ulem}
\newcommand*\circled[1]{\tikz[baseline=(char.base)]{
            \node[shape=circle,draw,inner sep=1pt] (char) {#1};}}

\newcolumntype{\$}{>{\global\let\currentrowstyle\relax}}
\newcolumntype{^}{>{\currentrowstyle}}

\algrenewcommand\algorithmicrequire{\textbf{Precondition:}}  
\algrenewcommand\algorithmicensure{\textbf{Postcondition:}}
\algnewcommand{\Inputs}[1]{%
	\State \textbf{Inputs:}
	\Statex \hspace*{\algorithmicindent}\parbox[t]{.8\linewidth}{\raggedright #1}
}
\algnewcommand{\Initialize}[1]{%
	\State \textbf{Initialize:}
	\Statex \hspace*{\algorithmicindent}\parbox[t]{.8\linewidth}{\raggedright #1}
}
\usepackage{environ}
\usepackage{soul}
\makeatletter
\newsavebox{\measure@tikzpicture}
\NewEnviron{scaletikzpicturetowidth}[1]{%
	\def\tikz@width{#1}%
	\begin{lrbox}{\measure@tikzpicture}%
		\BODY
	\end{lrbox}%
	\pgfmathparse{#1/\wd\measure@tikzpicture}%
	\BODY
}
\makeatother

\begin{document}
\pagestyle{empty}
%
\title{
LSTM-Based Proactive Congestion Management 
for 
Internet of Vehicle Networks}
%
%
%
\author{Aly Sabri Abdalla\textsuperscript{1}, Ahmad Al-Kabbany\textsuperscript{2}, Ehab F. Badran\textsuperscript{2}, and Vuk Marojevic\textsuperscript{1} \\
    \textsuperscript{1} Department of Electrical and Computer Engineering, Mississippi State University, MS 39762, USA \\ 
	\textsuperscript{2} Department of Electronics and Communications Engineering,
	Arab Academy for Science, Technology, and Maritime Transport, Alexandria, Egypt \\ 
}
\maketitle
\pagestyle{empty}
\begin{abstract}
Vehicle-to-everything (V2X) 
networks support a variety of safety, entertainment, and commercial applications. 
This is realized by applying the principles of 
the Internet of Vehicles (IoV) to facilitate connectivity among vehicles and between vehicles and roadside units (RSUs). Network congestion management 
is essential for 
IoVs 
and it represents a significant concern 
due to its impact on 
improving the efficiency of transportation systems and providing reliable 
communication among vehicles for the timely delivery of safety-critical packets. 
This paper introduces a framework for proactive congestion management for IoV networks. We generate congestion scenarios and a data set 
to predict the congestion 
using LSTM. We present the framework 
and the packet congestion dataset. Simulation results using SUMO with NS3 demonstrate the effectiveness of the framework for forecasting IoV network congestion 
and clustering/prioritizing packets employing 
recurrent neural networks.                 
\end{abstract}

\begin{IEEEkeywords}
V2X, IoV, Safety, Congestion Management, Machine Learning, NS3, SUMO, LSTM.
\end{IEEEkeywords}

\pagenumbering{gobble}

%
\IEEEpeerreviewmaketitle

\section{Introduction}
\textcolor{black}{
Vehicle-to-everything (V2X) communication is a new generation of wireless technology that enables vehicles to interact with the surrounding infrastructure and other mobile units on the road. The vehicles can communicate under the V2X umbrella in four modes: vehicle-to-vehicle (V2V), vehicle-to-infrastructure (V2I), vehicle-to-network (V2N), and vehicle-to-pedestrian (V2P) \cite{5G_V2X}. 
These modes of communication 
offer cooperative awareness to provide intelligent transportation services. 
There are two major V2X standards: Dedicated Short Range Communications (DSRC) and cellular-based V2X (C-V2X). DSRC is standardized by the IEEE, whereas C-V2X is defined by the Third Generation Partnership Project (3GPP) in the context of 4G and 5G networks~\cite{Amr2018VTC}.} 

\textcolor{black}{
The Internet of Vehicles (IoV) 
emerged for linking vehicles together through networks of interconnected 
units that communicate with one another and with the surrounding infrastructure. The IoV has the potential to reduce congestion, improve traffic flow, reduce road accidents, and improve overall transportation safety and efficiency through 
connectivity enabling real-time situational awareness.} 
\textcolor{black}{This can be done by exchanging safety-related information through the basic safety message (BSM)}. \textcolor{black}{The BSM is 
broadcast by a DSRC or C-V2X device to inform other nearby vehicles about vehicle position, speed, and so forth, as well as any circumstance that has been detected on the road. These messages are short, single-message transmissions that are sent regularly by all 
vehicles
~\cite{DSRC}.} 
Mechanisms to detect, avoid, and manage 
packets congestion 
to alleviate vehicle congestion 
is of critical importance \textcolor{black}{as mostly the road traffic congestion is 
leading to packet congestion via the increased communication demand within close proximity of vehicles. } 
\textcolor{black}{
\textcolor{black}{V2X congestion management mechanisms can be categorized into five classes: rate-based, 
power-based, 
carrier sense multiple access (CSMA)/collision avoidance (CA)-based, 
priority and scheduling-based, 
and hybrid~\cite{balador2022survey}. Several work items have discussed this problem in the literature~\cite{al2024enhancing, du2017stacked, elangovan2023accumulative, nithya2022cnn, zhang2023hybrid}. In~\cite{al2024enhancing}, the authors aim to anticipate road hazards for V2X via Long Short-Term Memory (LSTM) to enhance road safety. The authors of~\cite{du2017stacked} proposed stacked LSTM for improving the traffic conditions of vehicular networks. The work presented in~\cite{elangovan2023accumulative} introduces an accumulative approach to time series prediction based on LSTM, aiming to improve the accuracy of predictions. The study in~\cite{nithya2022cnn}   integrates convolutional neural network (CNN) and LSTM models to classify various Transmission Control Protocol (TCP) congestion control algorithms, enhancing the understanding and management of V2X congestion control. In~\cite{zhang2023hybrid}, the authors studied a hybrid congestion control for allocating traffic by employing a fusion model for traffic prediction.}   
} 
\textcolor{black}{While these work items have explored the use of LSTM for congestion control for vehicular networks, the integration of LSTM for congestion protection with the packet prioritization for proactive congestion management to improve delay and minimize dropped safety packets has not been well addressed. }
Therefore, this work introduces a proactive open-loop prioritized scheduling scheme that predicts IoV channel congestion to avoid such conditions. Using a LSTM for real-time forecasting and a historical congestion database with selective features, the proposed method ensures accurate prediction results. It classifies packets into two categories, ensuring high-priority safety messages are transmitted without delay even in congested scenarios. The key contributions of this paper are detailed further
\begin{itemize}
\item We design and implement a real-time framework for proactive network congestion management for IoV. 
Congestion events are predicted before they occur to facilitate packet scheduling actions that enhance the 
delivery ratio of the most important packets.
\item Through the course of this study, an enormous dataset 
is generated using Network Simulator Version~3 (NS3), which contains the congestion channel parameters to be used as historical entries for the prediction phase. 
\item 
A 
LSTM architecture is designed 
to maximize the accuracy of the prediction. 
\item 
The K-means clustering technique is employed to cluster packets into two classes and minimize 
\textcolor{black}{transmission} delay of safety-critical packets in congested environments. 
\end{itemize}
The structure of the paper is given as: 
Section II discusses the proposed multi-tier strategy for congestion prediction and packet classification. 
Section III numerically analyzes the performance. 
Section IV summarizes the main takeaways and Section V provides the concluding remarks. 
\section{PROPOSED SOLUTION}
The increasing population and expanding metropolitan areas has led to increased road congestion. 
Vehicular traffic congestion can lead to network congestion which severely impacts the network performance. 
Furthermore, the 
degradation of link reliability and packet delivery ratio due to a congested channel can have dire road safety consequences. 
As a result, it is critical to be able to handle 
congested situations in vehicular networks 
and provide the highest safety levels for vehicles and pedestrians. 
The problem focuses on developing an effective framework for channel congestion prediction enabling real-time adjustments that are protocol agnostic and scalable. 

\textcolor{black}{We introduce a framework to address real-time congestion prediction and packet classification in IoV networks. We utilize LSTM, known for its capability to capture temporal dependencies in traffic data, to predict network congestion. We implement stacked LSTM allowing the model to learn hierarchical traffic patterns through multiple LSTM layers to improve prediction accuracy. Additionally, we introduce K-means clustering for packet classification based on key features extracted from simulation data, enabling dynamic prioritization and timely delivery of safety-critical messages.}
 \begin{figure}[t]	
	\centering
	\includegraphics[width=0.48\textwidth]{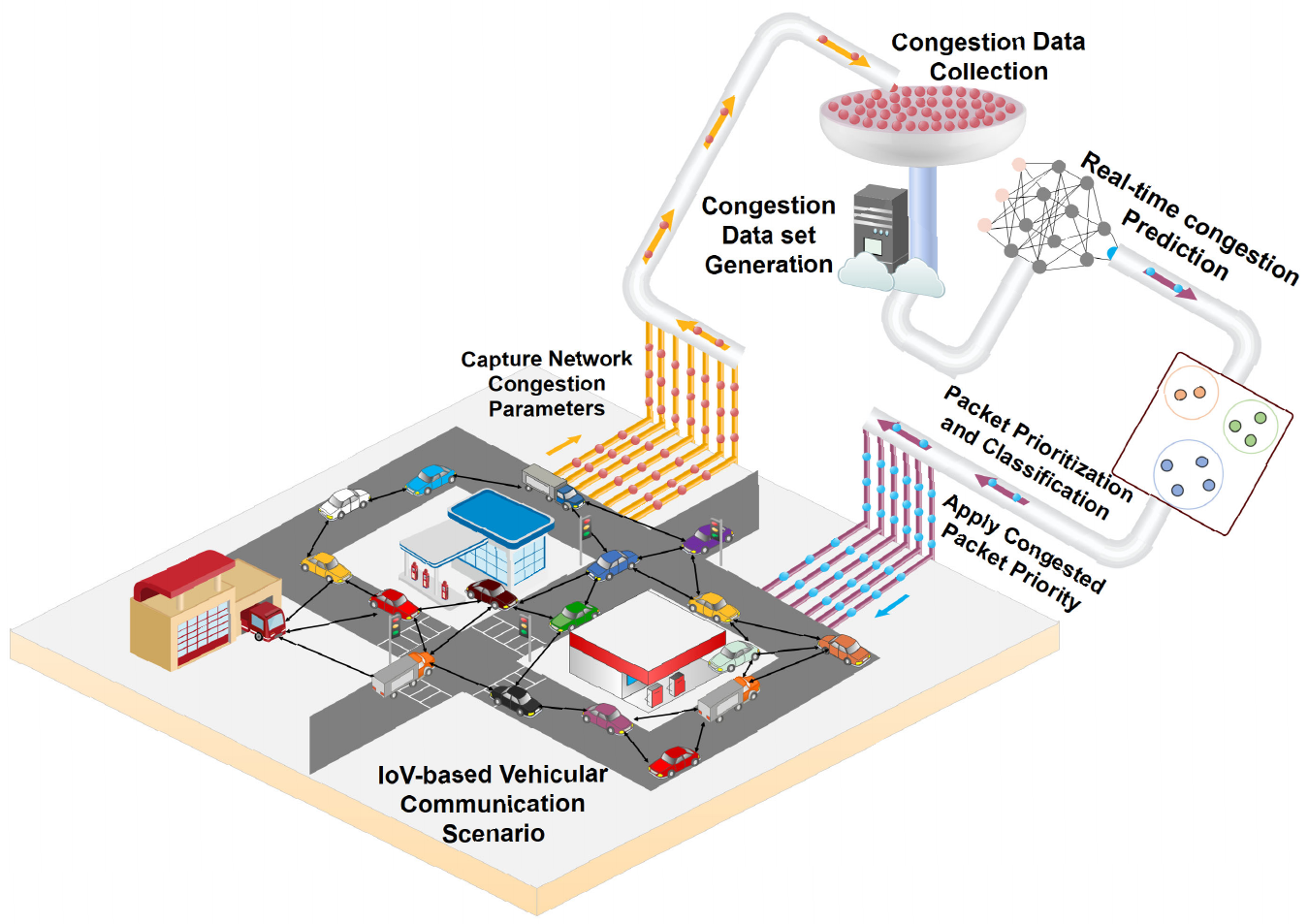}
	\captionsetup{justification=centering}
 \vspace{-5 pt}
	\caption{IoV-based vehicular networks congestion scenario with the proposed solution components.}	
 \label{V2XPropSol}
\end{figure}

Fig.~\ref{V2XPropSol} illustrates the main components of the proposed solution. The congestion management framework consists of three phases. Fig.~2 depicts the flow. 
First, 
the IoV-based vehicular communication network scenario is practically implemented and the congestion parameters are captured under different conditions to create 
data set. 
The generated data set is used as inputs to a neural network that predicts future crowded events in real time. Finally, the system assigns distinguished priorities depending on the content of the messages, which will assist the packet clustering during 
network congestion. 
 	



\subsection{Congestion data set generation}
The aim of this stage is to detect and store the congestion parameters of vehicles in the IoV-based vehicular network in a 
dataset. 
The generated 
dataset 
contains the number of dropped packets 
and the time of occurrence of these packet drops. 
\textcolor{black}{We are utilizing the NS3 as it incorporates realistic traffic models and mobility patterns that closely mimic real-world conditions. NS3 is widely recognized for testing V2X protocols because of its accuracy in emulating vehicular networks~\cite{ali20213gpp, drago2020millicar}.} The process is composed of three sub-processes: We first create an IoV-based vehicular communication scenario where the vehicles start transmitting packets under neutral congestion conditions. The IoV-based 
communication environment will be implemented through 
via multi-objective functions, such as initialization of grid parameters, creation of vehicles, and implementation of agents and traffic via a traffic simulator
as Simulation of Urban MObility (SUMO) tool. 

\textcolor{black}{The NS3 software applies a variable congestion rate which affects the 
transmission delay and packet delivery ratio of the system. The variable congestion rate is employed in our model by using a Markov chain, which can be referred to as the multi-state Markov chain error model. It is implemented to show the transition probabilities of congestion rates in matrix format where the rows reflect the current state and columns the next state of congestion, where these transitions are probabilistically determined, affecting transmission delay and packet delivery
ratio. 
Finally, for documentation purposes, the dataset measurements captured over the simulation time that is divided into a number of time slots within each time slot there will be a different congestion rate, also packet delivery ratio will be monitored and the number of drop packets will be counted, which makes the dataset more representative of real-world situations.} Fig.~\ref{V2XPropSol} shows the key components of the proposed solution including developing the IoV-based vehicular congestion dataset, real-time congestion prediction, and packet prioritization and classification. 
NS3 includes some functions to terminate and avoid congestion in the network, 
such as adaptive congestion window size and back-off timer which controls the retransmission timer period. 

This paper's focus is on designing multiple functions for monitoring several factors, which give the direct indication on the number of dropped packets during the simulation time. The main modified function 
to generate the dataset is the Congestion Action Difference (CongActDiff) counter. CongActDiff counts how many congestion flags were activated during a predefined interval within the overall simulation scenario. The congestion flag is enabled or activated when NS3 starts the process of changing the congestion window phase from a slow start to a fast retransmission phase. Moreover, modification of the congestion window phase occurs when the size of the congestion window becomes higher than 
a threshold 
as an indicator of the 
increase of the number of retransmitted packets. 
Algorithm 1 presents the congestion action function scenario. 
The simulation scenario is generated by executing the tool command language (TCL) script, which does the node initialization, variable congestion rate implementation, and congestion dataset entry storage. The observation of the final results can be done graphically by plotting CongActDiff over the simulation time 
or by checking the saved dataset. \textcolor{black}{This dataset consists of 25 records where each record is captured over 18 hours and covers 12,961,249 observations. 
} 
\vspace{-10 pt}
\begin{algorithm}
\footnotesize
	\caption{CongActDiff Algorithm}
	\begin{algorithmic}[1]
		\Require $ Allocate \ Packet \ P \ at \ transmitter \ side $
		\Initialize{$CongAct, \ CongInit,\ Packs,\ $\linebreak$CongActTime,\ and\ CongDiff  = 0$\linebreak$PackTH  = 350$\linebreak }
		\vspace{-5 pt}\For{$ each \ P \ in\ Packets $}
			\If{$ congestion\ flag[P]\gets \textsf{true}\ \&\& $ \\ \hspace{30 pt}$P$ $\neq$  $\ re-transmitting \ packet $}
				\State $ CongAct \leftarrow CongAct\ +\ 1 $
				\State $ CongActTime \gets instance()\ .\ clock()  $
				\State $ congestion\ flag \gets \textsf{false} $ 
    		\EndIf
    		\State $ Packs \leftarrow Packs\ +\ 1 $
    		\If{$ Packs \ counter = \ PackTH $}
    			\State $CongDiff = CongAct-CongInit$
    			\State $ CongInit\ = \ CongAct  $
    		\EndIf	
    	\EndFor	\\
    	\Return $CongDiff\ , \ CongActTime $
	\end{algorithmic}
\end{algorithm}
\vspace{-25 pt}
\subsection{Real-Time Congestion Prediction}
The prediction phase 
reiles on two elements to enable 
proactive actions in real-time: 
Firstly, the previous crowdedness log from networked vehicles, which contains the CongActDiff counter and the occurrence time during the day, is obtained. Secondly, the historical congestion dataset is fitted to 
the LSTM~\cite{LSTM}. In this paper, the aim is to reach the optimum structure of the LSTM network to achieve accurate real-time prediction readings from many perspectives. 
\textcolor{black}{LSTMs are particularly well-suited for handling time-series data generated by vehicular networks, as they can capture long-term dependencies and complex temporal patterns. The LSTMs, with forget, input, and output gates, address the vanishing and exploding gradient problems, ensuring stable training. Additionally, stacked LSTM layers enhance the model's capacity to learn hierarchical features, leading to improved real-time prediction accuracy. This capability is essential for proactive congestion management in IoV environments. Recent studies have demonstrated the effectiveness of LSTMs in similar applications~\cite{al2024enhancing, du2017stacked, elangovan2023accumulative, nithya2022cnn, zhang2023hybrid}}

LSTM models 
can be classified as a function of 
the batch size, the number of epochs, and the hidden layers. 
An LSTM with multiple hidden layers 
is referred to as a stacked LSTM. The stacked LSTM model can be seen as a representational optimization, where the added layers are used to defragment the learned features of prior layers and generate 
high-level representations. 
The second concern of the LSTM model structure is the size of the batch, which controls the update of the model weights. 
Because real-time congestion forecasting needs an enormous historical log of previous crowded 
events, which cannot be processed by the LSTM model at once, the dataset 
is divided into small groups called batches. The batch size controls the tradeoff 
between the training speed and the learning process. 
The third feature of the LSTM model 
is the effect of 
the number of epochs on the real-time prediction process. In particular, when the model is exposed to the same random samples of the dataset many times, the model may memorize it. As a result, the LSTM network becomes over-fitted. Too few numbers of epochs, on the other hand, may lead to under-fitting. 
The appropriate choices for the epoch and batch sizes is presented in the next section. 
This solution for the congestion prediction problem is classified as an LSTM sequence-to-sequence $(seq2seq)$ prediction, 
where the input and output are time series sequences.
 \begin{figure*}[h!]
	\centering
	\includegraphics[width=0.77\textwidth]{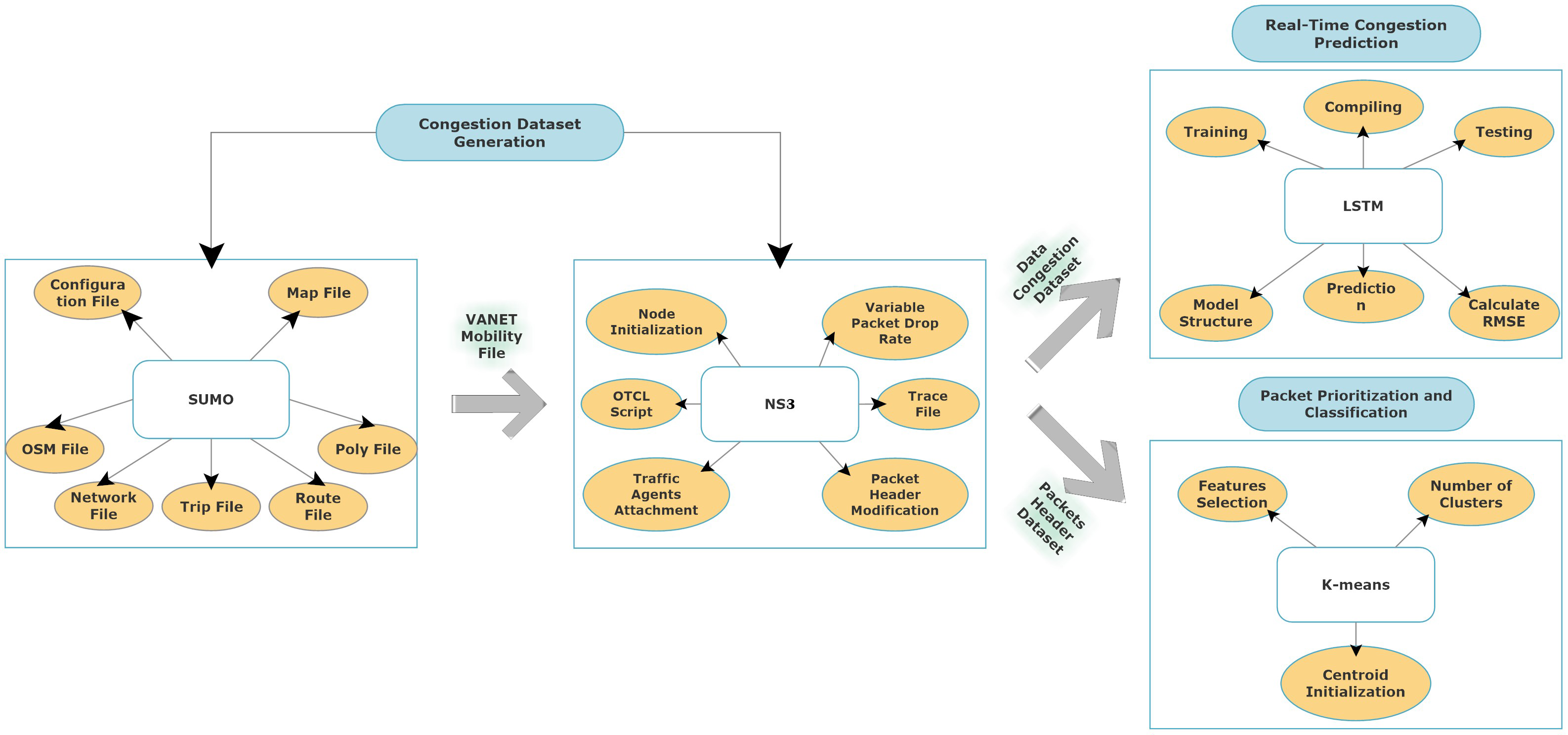}
 \vspace{-10 pt}
	\caption{Overall process of proposed strategy. 
 }
	\label{total}
  \vspace{-10 pt}
\end{figure*}    
\begin{figure}[h]	
\vspace{-12 pt}
	\centering
	\includegraphics[width=0.7\columnwidth]{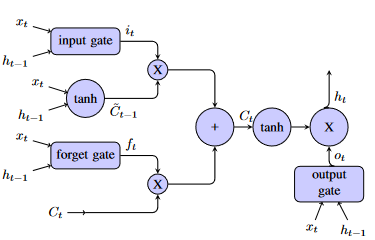}
 \vspace{-6 pt}
	\caption{\textcolor{black}{ LSTM \emph{Memory Block} structure.}}	
 \vspace{-19 pt}
\end{figure}
\textcolor{black}{Fig.~3 shows the internal structure of LSTM \emph{Memory Block}. 
We denote $C_t$ and $h_t$ as the cell state and final hidden state at time $t$, respectively. 
We consider $x_t$ the vector of the input sequence at time step $t$. Eq.~\ref{1} to Eq.~\ref{6} shows how the single LSTM works, while Eq.~\ref{7} generalizes this to the stacked LSTM~\cite{du2017stacked}. First, the computation of the forget and input gate vectors are given as follows: 
\vspace{-7 pt}
\begin{equation}
\vspace{-5 pt}
    f_{t}  = \sigma(W^{f}_{1}.x_{t}+W^{f}_{h}.h_{t-1}+b_{f}), \label{1}
\tag{1}
\end{equation}
\begin{equation}
i_{t}  = \sigma(W^{i}_{1}.x_{t}+W^{i}_{h}.h_{t-1}+b_{i}),\label{2}
\tag{2}
\vspace{-7 pt}
\end{equation}
Then, we compute a vector of new candidate values that can be added to the cell state: 
\vspace{-7 pt}
\begin{equation}
\vspace{-5 pt}
\tilde{C}_{t}  = tanh(W^{C}_{1}.x_{t}+W^{C}_{h}.h_{t-1}+b_{C}),\label{3}
\tag{3}
\vspace{-3 pt}
\end{equation}
The next step is to update the new cell status $C_t$ with the new information and the calculated gate vectors:
\vspace{-7 pt}
\begin{equation}
{C}_{t}  = i_{t}* \tilde{C}_{t}+f_{t}*C_{t-1}, \label{4}
\tag{4}
\vspace{-6 pt}
\end{equation} 
Then, the output gate determines the required portion  of the cell state will be at the output:
\vspace{-7 pt}
\begin{equation}
o_{t}  = \sigma(W^{o}_{1}.x_{t}+W^{o}_{h}.h_{t-1}+b_{o}),\label{5}
\tag{5}
\vspace{-6 pt}
\end{equation}
Finally, we compute the hidden state at the output using the output gate vector
and the cell state:
\vspace{-7 pt}
\begin{equation}
{h}_{t}  = o_{t}* tanh({C}_{t})   \label{6}
\tag{6}
\vspace{-6 pt}
\end{equation}
The connection between the input $x_{t}$ and three LSTM gates and cell input is controlled by weight matrices $W^{i}_{1}, W^{f}_{1}, W^{o}_{1}$, and $W^{C}_{1}$, while $W^{i}_{h}, W^{f}_{h}, W^{o}_{h}$, and $W^{C}_{h}$ are weight matrices adjusting the link between previous hidden layer state $h_{t-1}$ and three gates plus cell input. The $b_{i}, b_{f}, b_{o}$, and $b_{C}$ are bias factors of input, forget, output, and cell input respectively, a sigmoid activation function used in LSTM calculation and $tanh$ expresses the hyperbolic tangent function which used to overcome the vanishing gradient problem. In a stacked LSTM, these operations are repeated for each $l$ layer:
\vspace{-5 pt}
\begin{equation}
h^l_t, C^l_t = \text{LSTM}(x_t, h^{l-1}_{t-1}, C^{l-1}_t)
\label{7}
\tag{7}
\vspace{-8 pt}
\end{equation}
}

\vspace{-15 pt}
\subsection{Packet Prioritization and Classification} 
Congestion scenarios and channels can effectively decrease the reliability of the system. The previous crowdedness disadvantage has a damageable weight on the packet delivery ratio. 
The third procedure proposed in our system is crystallizing the importance of delivering safety message in congestion conditions. 
The predicted congestion facilitates system flexibility 
by giving higher priority to safety critical messages for improving their delivery ratio in congested conditions. 

Prioritization of packets is based on the content of the message by adding a priority field to the TCP header within the creation of packets in the Medium Access Control (MAC) layer before transmission. 
A classification technique is then used to classify the flow of packets that will face congestion in the channel based on the priority and time to live (TTL) fields in the packet header. The TTL field defines the remaining hops before the intermediate node. If any node at any time 
receives a packet with TTL equaling zero then this packet will be dropped to avoid formulating loops inside the network. 
By finding the almost expired and highest priority safety message, this packet is transmitted with minimum hops and delays. 

The K-means unsupervised machine learning algorithm is implemented 
for clustering packets~\cite{Kmeans}. It is 
a widely used clustering solution because of its speed of convergence, efficient handling of large dimension features, and simplicity of implementation \cite{Kmeans}. 
K-means classifies packets upon training with the large dataset generated for header fields of packets during the simulation time. NS3 by default does not generate an output file containing any information about the packet header. While the TTL field is already defined as one of the packet header fields, the other classification feature (priority) needs to be added to the header structure of NS3. 
The dataset containing the modified packet header fields after the NS3 modifications can then be fed to the K-means. 

\textcolor{black}{The K-means algorithm is utilized for classifying packets based on features extracted from simulation data, such as packet size, TTL, priority, generation time, and hop count. This algorithm is selected for its efficiency and simplicity in handling large datasets. Initially, centroids are chosen either randomly or using the K-means++ method to ensure optimal starting points. Each packet is then assigned to the nearest centroid based on Euclidean distance, forming clusters. The centroids are iteratively updated by calculating the mean of the packets in each cluster until convergence is achieved. The performance of the clustering is validated using the Silhouette Score, ensuring well-defined and distinct clusters. This approach enables dynamic prioritization and timely delivery of safety-critical messages, enhancing the overall effectiveness of the congestion management framework.}

The new packet header trace 
contains 
header fields, such as source IP, destination IP, TTL, packet size, and recently added field priority. \textcolor{black}{Algorithm 2 indicates the steps needed to add the new priority field to the IP header of the generated packets at the transmitter. 
Without loss of generality, the condition 
that define safety packets used in our algorithm is that the size is below $100$ bytes. Furthermore, the defined safety message will take random priority between 1~to~10, on the other hand, normal packets will have priority from 11~to~20.}
\vspace{-9 pt}
\begin{algorithm}
\color{black}
\footnotesize
	\caption{\color{black}\small Priority Field Algorithm}
	\begin{algorithmic}[1]
		\Require $ Allocate \ Packet \ P \ at \ transmitter \ side $
		\For{$ each \ P \ in\ Packets $}
		\State $PH \gets hdr\ \_\ access(p)$
		\If{$ PH \rightarrow ptype(P)\gets Routing\ $}
		\If{$ PH \rightarrow size(P) <= \ 100  $}
		\State $PH \rightarrow prio(P)= random(1~10)$
		\vspace{-2 pt}\Else\vspace{-2 pt}
		\State $PH \rightarrow prio(P)= random(11~20)$
		\vspace{-2 pt}\EndIf
		\vspace{-2 pt}\EndIf			
		\vspace{-2 pt}\EndFor
	\end{algorithmic}
\end{algorithm}
\vspace{-15 pt}

\section{PERFORMANCE EVALUATION}
\textcolor{black}{ 
Fig.~\ref{total} shows 
the 
stages of the proposed solution and the 
operations that occur in each of the respective stages. This framework is implemented employing different software tools including SUMO, NS3, LSTM, and K-Means. 
The proposed framework executes in a centralized node in an IoV network and can be integrated into radio access network (RAN) intelligent controllers of future Open-RAN systems~\cite{AlyORAN} or in the form of O-RAN distributed learning~\cite{10520642}. 
}
\vspace{-5 pt}
\subsection{
Simulation Parameters}
SUMO is used to generate mobility patterns of vehicles \cite{6}. The scenario of this paper is based on a real map extracted from OpenStreetMap (OSM) for the Arab Academy for Science and Technology (AAST) campus in Alexandria, Egypt.  
The simulation area is two dimensional $1530$ m length and $1052$ m width with $93$ junctions and $134$ single lane route.
%
\textcolor{black}{The simulation scenario employs the two-ray ground radio propagation model with omnidirectional antennas and the 802.11p MAC protocol for IoV 
communications. 
The packets are routed between vehicles using the ad hoc on-demand distance vector (AODV) routing protocol. The number of vehicles simulated in this scenario is 150 driving at a fixed speed of 25 kilometers per hour. The simulation time is 450 hours. }
We evaluate our framework for four congestion stages of $0.1$, $0.35$, $0.2$ and $0.3$ packet drop rates. 
The next stage is selected using the Markov chain matrix mechanism. 
\begin{table*}[!htbp]
	\renewrobustcmd{\bfseries}{\fontseries{b}\selectfont}
	\renewrobustcmd{\boldmath}{}
	\centering
 	\caption{Categorical LSTM model structure results.}
	\resizebox{\textwidth}{!}{%
		\begin{NiceTabular}{|l|l|c|c|c|c|c|c|}[name=R]
			\hline
			\multirow{10}{*}{LSTM Model Structure} &                                              & \multicolumn{1}{l|}{Batch Size} & \multicolumn{1}{l|}{No.of Epochs} & \begin{tabular}[c]{@{}c@{}}Hidden layers Size \\  (Nuerons)\end{tabular} & Training Accuracy                      & Test Accuracy     & \multicolumn{1}{l|}{Overall RMSE} \\ \cline{2-8} 
			& \multirow{3}{*}{Based on Batch Size}         & \bfseries 32                              &\bfseries 1000                              & \bfseries 1 layer (64)                                                               & \bfseries 98.23 & \bfseries{99.16}  & \bfseries 2454.07                           \\ \cline{3-8} 
			&                                              & 64                              & 1000                              & 1 layer (64)                                                               & 98.20                      & 98.59  & 4325.29                           \\ \cline{3-8} 
			&                                              & 128                             & 1000                              & 1 layer (64)                                                               & 98.51                      & 98.57 & 4388.65                           \\ \cline{2-8} 
			& \multirow{3}{*}{Based on No.of Epochs}       & 32                              & 350                               & 1 layer(64)                                                               & 97.15                      & 95.26 & 13712.17                          \\ \cline{3-8} 
			&                                              & 32                              & 750                               & 1 layer (64)                                                               & 98.56                      & 96.62 & 10368.44                          \\ \cline{3-8} 
			&                                              &\bfseries 32                              & \bfseries 1000                              &\bfseries 1 layer (64)                                                               &\bfseries 98.23                      &\bfseries 99.16  &\bfseries 2454.07                           \\ \cline{2-8} 
			& \multirow{3}{*}{Based on Hidden Layers Size} & \color{black}64                              & \color{black}1000                              & \color{black}1 layer(64)                                                                & \color{black}98.20                      & \color{black}98.59  & \color{black}4325.29                           \\ \cline{3-8} 
			&                                              & 64                              & 1000                              & 2 layers(64-32)                                                            & 98.40                      & 98.86 & 3497.03                           \\ \cline{3-8} 
			&                                              & \color{black}\bfseries 64                              & \color{black}\bfseries 1000                              & \color{black}\bfseries 2 layers(64-16)                                                            &\color{black}\bfseries 98.53                      & \color{black}\bfseries 99.26 & \color{black}\bfseries 2147.31                           \\ \hline
		\end{NiceTabular}
	}
\end{table*}
              
%

\vspace{-5 pt}
\subsection{LSTM 
Network Structure and Results}
The 
LSTM 
is fed with the generated dataset from the previous steps. 
The data set is partitioned as $70\%$ and $30\%$ for training and testing. 
We consider different LSTM 
structures based on the size of the batch, number of epochs and the number of hidden layers. Table I shows the obtained training and testing accuracy and the overall root mean square error (RMSE) of the different LSTM models based on the previously mentioned categories. 
The results show that the performance of the LSTM model greatly improves by decreasing the size of the batch from $128$ to $32$. The small batch size is capable of effectively capturing the generalization pattern of the dataset. The second parameter used to investigate the LSTM model structure is the number of epochs, where more epochs enhance the accuracy of predictions as shown in Table~1. 

The depth of the 
the LSTM model refers to the number of hidden layers and the number of neurons 
in each layer. Table I includes the results obtained for different depths of the network. Where we can notice that using two or more layers provides an accurate understanding of 
the complex representations in the dataset. 
Also, the use of a small number of neurons in the hidden layer can lead to 
under-fitting.\begin{figure}[h]	
\vspace{-12 pt}
	\centering
	\includegraphics[width=0.75\columnwidth]{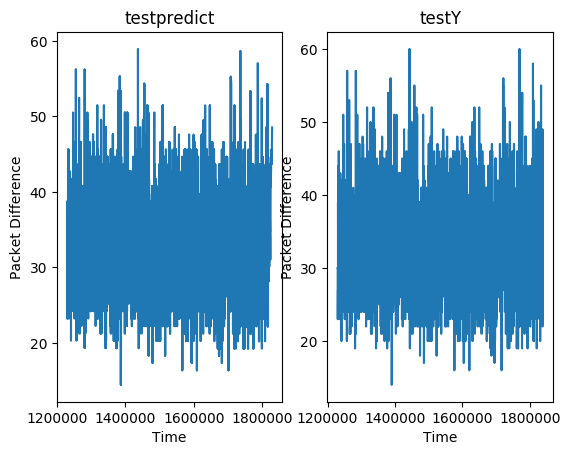}
 \vspace{-6 pt}
	\caption{Predicted test verus test data.}	
 \vspace{-12 pt}
\end{figure} On the other hand, a large number of neurons per hidden layer will be the reason for an over-fitted 
model. 
After many trials 
the proposed model 
is composed of two hidden layers with $64$ and $16$ neurons respectively, a batch size of $64$ samples, and $1000$ epochs. The RMSE is added to evaluate the accuracy of results. 
The proposed model structure achieves a test   
accuracy equivalent to $99.3\%$ with $2147$ RMSE. 

Fig.~4 contains samples taken during the simulation from the predicted 
and the actual test data to illustrate the efficiency of the LSTM model. \textcolor{black}{
Table 1 shows that the two-layer stacked LSTM significantly outperforms the single LSTM. Specifically, the two-layer stacked LSTM achieves an RMSE that is less than 50\% of the RMSE of the single LSTM, with a prediction accuracy of 98.53\%. This high level of accuracy, combined with efficient training, highlights the stacked LSTM's ability to capture long-term dependencies, making it ideal for real-time congestion prediction in IoV networks.} 

\begin{figure}[h]	
\vspace{-12 pt}
	\centering
	\includegraphics[width=0.7\columnwidth]{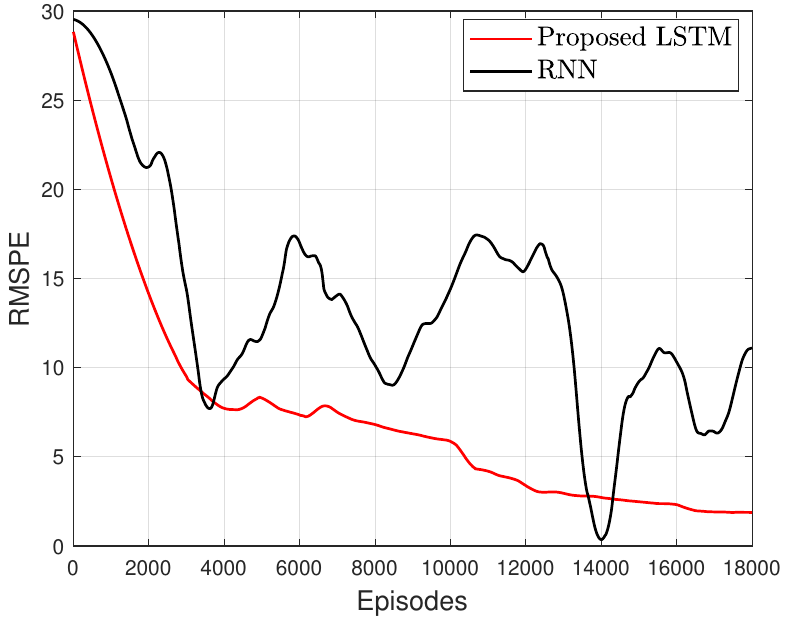}
 \vspace{-6 pt}
	\caption{\textcolor{black}{The RMSPE performance of the proposed LSTM solution versus RNN benchmark technique.}}	
 \vspace{-8 pt}
\end{figure}

\textcolor{black}{Fig.~5 illustrates the Root Mean Square Percentage Error (RMSPE) comparison between the proposed LSTM-based method and a standard Recurrent Neural Network (RNN) over the training episodes. The RMSPE, a metric used to evaluate the accuracy of predictive models, is defined as: $\textcolor{black}{RMSPE= \sqrt{\frac{1}{n}\sum^n_{i=1}\left(\frac{y_i-\hat{y_i}}{y_i} \right)^2} \times 100}$, } 
\textcolor{black}{where $y_i$ is the actual value, $\hat{y_i}$ is the predicted value, and $n$ is the number of observations. The proposed LSTM method demonstrates superior performance, rapidly converging to a lower RMSPE within the initial episodes and maintaining a stable error rate of approximately 3-5\%. In contrast, the RNN shows significant fluctuations with a higher RMSPE, stabilizing around 8-12\%. This outcome highlights the LSTM's ability to effectively capture long-term dependencies and overcome the vanishing gradient problem that degrades the RNN performance, resulting in more accurate and stable predictions.
}
\subsection{Packet Prioritization and Classification Model}
The K-means classification model is fed by a packet header dataset during the simulation time, which contains $1,048,576$ 
entries. This 
dataset contains all previously mentioned packet header fields despite that the effective fields which are used to categorize the packet are the TTL and the priority. Therefore, based on the provided classification features the aim of this step is to classify all created packets into two classes, the most critical and the least critical packets. 
The most critical packets have a small TTL and high priority and will 
be sent first and immediately. 
\begin{figure}[h]	
\vspace{-18 pt}
	\centering
	\includegraphics[width=0.73\columnwidth]{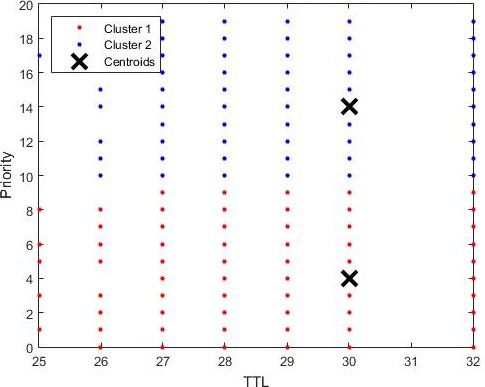}
 \vspace{-6 pt}
	\caption{K-means cluster assignments and centroids.}
 \label{Kmeans}
 \vspace{-8 pt}
\end{figure}
The least critical packets deal with the channel's normal conditions, which 
are sent 
in case of free congestion or dropped channel. 
For the K-means++ implementation presented in Fig.~\ref{Kmeans}, the candidate center 
must minimize the sum squared distances from each case point being clustered to the first randomly chosen cluster center. 

\section{Takeaways}

\textcolor{black}{Building a rich dataset for our IoV 
research 
in this paper is challenging for these reasons: the need to collect data from multiple sources, 
the need to capture data at different levels of granularity, 
and the sheer amount of data needed. 
The quality of the 
data is a fundamental aspect of data-driven research and has an effect on the accuracy of the models and inferences. 
Data quality is defined in terms of validity, reliability, 
and representativeness. 
The quality level can be measured using the fidelity of the data  
via how closely the data mimics the desired characteristics of the real world by measuring the accuracy, completeness, timeliness, relevance, and authenticity of the data. 
Defining accurate and consistent data acquisition procedures is essential and 
we have performed several validation checks, including \circled{1} checking for dataset completeness by ensuring 
the availability of necessary data points and the 
lack of significant gaps or missing values, \circled{2} checking the data for errors and inconsistencies to verify 
its accuracy, and 
\circled{3} checking for any potential biases in the data that could impact the results or conclusions drawn from the data. \textcolor{black}{In addition, we aim to do more verification of the dataset in the future by comparing the data points with actual data measured in real-world scenarios performed over V2X testbeds/industrial test tracks or in production IoV environments during trials similar to what we have done for 
aerial vehicle research~\cite{9625086}.} }


\section{Conclusion}
In this research, we are concerned with the problem of network congestion management 
for the IoV. Different from previous techniques in the literature, we have proposed a novel pipeline that features a proactive congestion prediction 
using LSTM 
to mitigate packet loss and ensure reliable safety message delivery. 
We adopt 
K-means 
clustering to classify packets based on priority and TTL parameters. 
Performance evaluation for different LSTM structures and training hyperparamters are presented. 
The adopted LSTM architecture 
performance exceeds $98\%$ prediction accuracy, enhancing delivery ratio and reducing delays. 
We were able to cluster the packets effectively into two groups of high and low message priority levels. 
\vspace{-10 pt}

%

\appendices

\ifCLASSOPTIONcaptionsoff
  \newpage
\fi



%
\section*{\textcolor{black}{Acknowledgment}}
\vspace{-1 pt}
\noindent
The work performed by Aly S. Abdalla and Vuk Marojevic was in part supported by NSF award CNS-
2120442. 
\bibliographystyle{IEEEtran}
\bibliography{ref}

\begin{thebibliography}{10}
\providecommand{\url}[1]{#1}
\csname url@samestyle\endcsname
\providecommand{\newblock}{\relax}
\providecommand{\bibinfo}[2]{#2}
\providecommand{\BIBentrySTDinterwordspacing}{\spaceskip=0pt\relax}
\providecommand{\BIBentryALTinterwordstretchfactor}{4}
\providecommand{\BIBentryALTinterwordspacing}{\spaceskip=\fontdimen2\font plus
\BIBentryALTinterwordstretchfactor\fontdimen3\font minus \fontdimen4\font\relax}
\providecommand{\BIBforeignlanguage}[2]{{%
\expandafter\ifx\csname l@#1\endcsname\relax
\typeout{** WARNING: IEEEtran.bst: No hyphenation pattern has been}%
\typeout{** loaded for the language `#1'. Using the pattern for}%
\typeout{** the default language instead.}%
\else
\language=\csname l@#1\endcsname
\fi
#2}}
\providecommand{\BIBdecl}{\relax}
\BIBdecl

\bibitem{5G_V2X}
A.~Alalewi \emph{et~al.}, ``{On 5G-V2X Use Cases and Enabling Technologies: A Comprehensive Survey},'' \emph{IEEE Access}, vol.~9, pp. 10--37, 2021.

\bibitem{Amr2018VTC}
A.~Nabil, K.~Kaur, C.~Dietrich, and V.~Marojevic, ``Performance analysis of sensing-based semi-persistent scheduling in c-v2x networks,'' in \emph{2018 IEEE 88th Vehicular Technology Conference (VTC-Fall)}, 2018, pp. 1--5.

\bibitem{DSRC}
K.~Z. Ghafoor \emph{et~al.}, ``{Enabling Efficient Coexistence of DSRC and C-V2X in Vehicular Networks},'' \emph{IEEE Wireless Communications}, vol.~27, no.~2, pp. 134--140, 2020.

\bibitem{balador2022survey}
A.~Balador \emph{et~al.}, ``{Survey on decentralized congestion control methods for vehicular communication},'' \emph{Vehicular Communications}, vol.~33, p. 100394, 2022.

\bibitem{al2024enhancing}
F.~A. Al-Yarimi, ``{Enhancing road safety through advanced predictive analytics in V2X communication networks},'' \emph{Computers and Electrical Engineering}, vol. 115, p. 109134, 2024.

\bibitem{du2017stacked}
X.~Du \emph{et~al.}, ``{Stacked LSTM deep learning model for traffic prediction in vehicle-to-vehicle communication},'' in \emph{2017 IEEE 86th Vehicular Technology Conference (VTC-Fall)}.\hskip 1em plus 0.5em minus 0.4em\relax IEEE, 2017, pp. 1--5.

\bibitem{elangovan2023accumulative}
V.~Elangovan \emph{et~al.}, ``{An Accumulative Method to Time Series Prediction for Vehicle Communication},'' in \emph{2023 IEEE Vehicle Power and Propulsion Conference (VPPC)}.\hskip 1em plus 0.5em minus 0.4em\relax IEEE, 2023, pp. 1--6.

\bibitem{nithya2022cnn}
B.~Nithya \emph{et~al.}, ``{A CNN-LSTM Approach for Classification of Major TCP Congestion Control Algorithms},'' in \emph{Intelligent Sustainable Systems: Selected Papers of WorldS4 2021, Volume 2}.\hskip 1em plus 0.5em minus 0.4em\relax Springer, 2022, pp. 579--591.

\bibitem{zhang2023hybrid}
T.~Zhang \emph{et~al.}, ``{A hybrid method of traffic congestion prediction and control},'' \emph{IEEE Access}, vol.~11, pp. 36\,471--36\,491, 2023.

\bibitem{ali20213gpp}
Z.~Ali \emph{et~al.}, ``{3GPP NR V2X mode 2: Overview, models and system-level evaluation},'' \emph{IEEE Access}, vol.~9, pp. 89\,554--89\,579, 2021.

\bibitem{drago2020millicar}
M.~Drago \emph{et~al.}, ``{MilliCar: An ns-3 module for mmWave NR V2X networks},'' in \emph{Proceedings of the 2020 Workshop on ns-3}, 2020, pp. 9--16.

\bibitem{LSTM}
F.~Karim \emph{et~al.}, ``{LSTM Fully Convolutional Networks for Time Series Classification},'' \emph{IEEE Access}, vol.~6, pp. 1662--1669, 2018.

\bibitem{Kmeans}
K.~P. Sinaga and M.-S. Yang, ``{Unsupervised K-Means Clustering Algorithm},'' \emph{IEEE Access}, vol.~8, pp. 80\,716--80\,727, 2020.

\bibitem{AlyORAN}
A.~S. Abdalla \emph{et~al.}, ``{Toward Next Generation Open Radio Access Networks--What O-RAN Can and Cannot Do!}'' \emph{IEEE Network}, pp. 1--8, 2022.

\bibitem{10520642}
M.~Kouchaki \emph{et~al.}, ``{OpenAI dApp: An Open AI Platform for Distributed Federated Reinforcement Learning Apps in O-RAN},'' in \emph{2023 IEEE Future Networks World Forum (FNWF)}, 2023, pp. 1--6.

\bibitem{6}
M.~Behrisch \emph{et~al.}, ``Sumo - simulation of urban mobility: An overview,'' in \emph{in SIMUL 2011, The Third International Conference on Advances in System Simulation}, 2011, pp. 63--68.

\bibitem{9625086}
A.~S. Abdalla \emph{et~al.}, ``Open-source software radio performance for cellular communications research with uav users,'' in \emph{2021 IEEE 94th Vehicular Technology Conference (VTC2021-Fall)}, 2021, pp. 1--6.

\end{thebibliography}
\end{document}